\documentclass[%
 reprint,
 amsmath,amssymb,
 aps,
]{revtex4-2}
\usepackage{booktabs}
\usepackage{siunitx}
\usepackage{graphicx}
\usepackage{dcolumn}
\usepackage{bm}
\usepackage{braket}
\usepackage{color}

\begin{document}

\preprint{APS/123-QED}

\title{Raising the Cavity Frequency in cQED}
\author{Raymond A. Mencia}
\author{Taketo Imaizumi}
\author{Igor A. Golovchanskiy}
\author{Andrea Lizzit}
\author{Vladimir E. Manucharyan}%
\affiliation{%
 Institute of Physics, Ecole Polytechnique Federale de Lausanne, CH 1015
}%

\date{\today}

\begin{abstract} 

The basic element of circuit quantum electrodynamics (cQED) is a cavity resonator strongly coupled to a superconducting qubit. Since the inception of the field, the choice of the cavity frequency was, with a few exceptions, been limited to a narrow range around 7 GHz due to a variety of fundamental and practical considerations. Here we report the first cQED implementation, where the qubit remains a regular transmon at about 5 GHz frequency, but the cavity’s fundamental mode raises to 21 GHz. We demonstrate that (i) the dispersive shift remains in the conventional MHz range despite the large qubit-cavity detuning, (ii) the quantum efficiency of the qubit readout reaches 8\%, (iii) the qubit's energy relaxation quality factor exceeds $10^7$, (iv) the qubit coherence time reproducibly exceeds $100~\mu\rm{s}$ and can reach above $300~\mu\rm{s}$ with a single echoing $\pi$-pulse correction. The readout error is currently limited by an accidental resonant excitation of a non-computational state, the elimination of which requires minor adjustments to the device parameters. 
Nevertheless, we were able to initialize the qubit in a repeated measurement by post-selection with $2\times 10^{-3}$ error and achieve $4\times 10^{-3}$ state assignment error.
These results encourage in-depth explorations of potentially transformative advantages of high-frequency cavities without compromising existing qubit functionality. 

\end{abstract}

\maketitle

\section{Introduction}

The key to controlling quantum information in cQED is the dispersive interaction regime, where the qubit and the cavity are detuned enough in frequency to prevent energy exchange, but not excessively, as to retain a second-order (dispersive) shift in the cavity frequency conditioned on the qubit state \cite{blais2004cavity,blais2021circuit}. A vast majority of experiments involving high-coherence qubits implement the dispersive regime using qubits (typically transmons) operating at a frequency of about 5 GHz with cavities slightly detuned upward to about $6-8$ GHz range \cite{kono2024mechanically,tuokkola2025methods,bland20252d}. Can it be that such a two-decade-long arrangement is far from optimal?

One major consideration is the dephasing of qubits by residual thermal photons in the cavity \cite{sears2012photon, rigetti2012superconducting, zhang2017suppression, yeh2017microwave, yan2018distinguishing, wang2019cavity}. The effective cavity temperature usually gets stuck at about $40-50~\rm{mK}$, apparently due to the lack of thermalization of lossy elements in microwave attenuators, cables, and other components. Given the relation $h\times 20~\rm{GHz} \approx k_B \times 1~\rm{K}$, a $7~\rm{GHz}$ cavity overcoupled to a $50~\rm{mK}$ environment would have a mean residual photon number $\bar n \gtrsim 10^{-3}$. Raising the cavity frequency just two-fold would drop the Boltzmann exponent to $\bar n < 10^{-6}$, thus putting aside the problem of low-temperature thermalization \cite{pobell2013matter}. Furthermore, higher frequency cavities could enable the operation of superconducting qubits at intentionally elevated temperatures without the loss of coherence, potentially improving the scalability of quantum processors. In a recent work \cite{Anferov2024}, the transmon and the cavity frequencies were both raised above $20~\rm{GHz}$ with the qubit coherence time measured at temperatures above $200~\rm{mK}$. Such a setup demonstrated the expected insensitivity of dephasing time to temperature, but the low baseline coherence time of the high-frequency transmon ($T_2 < 1~\rm{\mu s}$) left the feasibility of higher-frequency cQED under question. 

A more subtle but crucial consideration is the readout error. It is now known that an off-resonant drive at the cavity frequency experienced by the transmon qubit during a readout operation induces multi-photon transitions to the non-computational states, thereby breaking the dispersive approximation and limiting the readout fidelity \cite{sank2016measurement, shillito2022dynamics, khezri2023measurement,dumas2024measurement,nesterov2024measurement, dai2025spectroscopy, connolly2025full}. Even heavily optimized measurement setups struggle to reduce the readout error far below $1\%$ \cite{jeffrey2014fast,walter2017rapid,bengtsson2024model,swiadek2024enhancing,sunada2024photon}, a major limitation for quantum error correction algorithms. In a very recent experiment \cite{kurilovich2025high}, though, it was shown that such drive-induced ``leakage” transitions can be suppressed to a large degree by choosing a very high qubit-cavity detuning, while retaining the dispersive interaction by appropriately increasing the qubit-cavity coupling. The cavity was pushed up in frequency only slightly, but the transmon frequency was lowered dramatically to a sub-GHz range, such that the ratio of the cavity and the qubit frequency reached about 12. Although the applicability of sub-GHz transmons may raise questions, this remarkable demonstration urges further explorations of, let us call it ``ultra-dispersive” regime of cQED.

Motivated by the recent experimental results of Ref.~\cite{Anferov2024,kurilovich2025high}, we refine our earlier question to: would it be feasible to raise the cavity frequency in cQED without changing the transmon qubit? In particular, would we be able to sustain the microwave hygiene of qubits facing an unconventional cavity environment, and hence keep the coherence time at the state-of-the-art level? Would we be able to efficiently collect readout photons carrying information on the qubit state in the unexplored frequency range, where the effects of impedance mismatch and losses are magnified? Is it even possible to just raise the cavity frequency without losing the dispersive interaction? In this paper, we positively resolve the above prerequisites for a 21~GHz cavity. The studies of leakage and temperature resilience are reserved for future work, but here we already report promising results. First, the leakage from the ground state is suppressed to a degree enabling us to initialize the qubit by a repeated measurement with an error of about $2\times 10^{-3}$. Second, the cavity's residual photon population was constrained to be in the several $10^{-4}$ range or lower despite the absence of heavy filtering of the measurement lines. Our demonstrations build a foundation for an imminent migration of cQED experiments into the microwave K-band (18-26.5 GHz).

\begin{figure*}[ht]
  \centering
  \includegraphics[width=16cm]{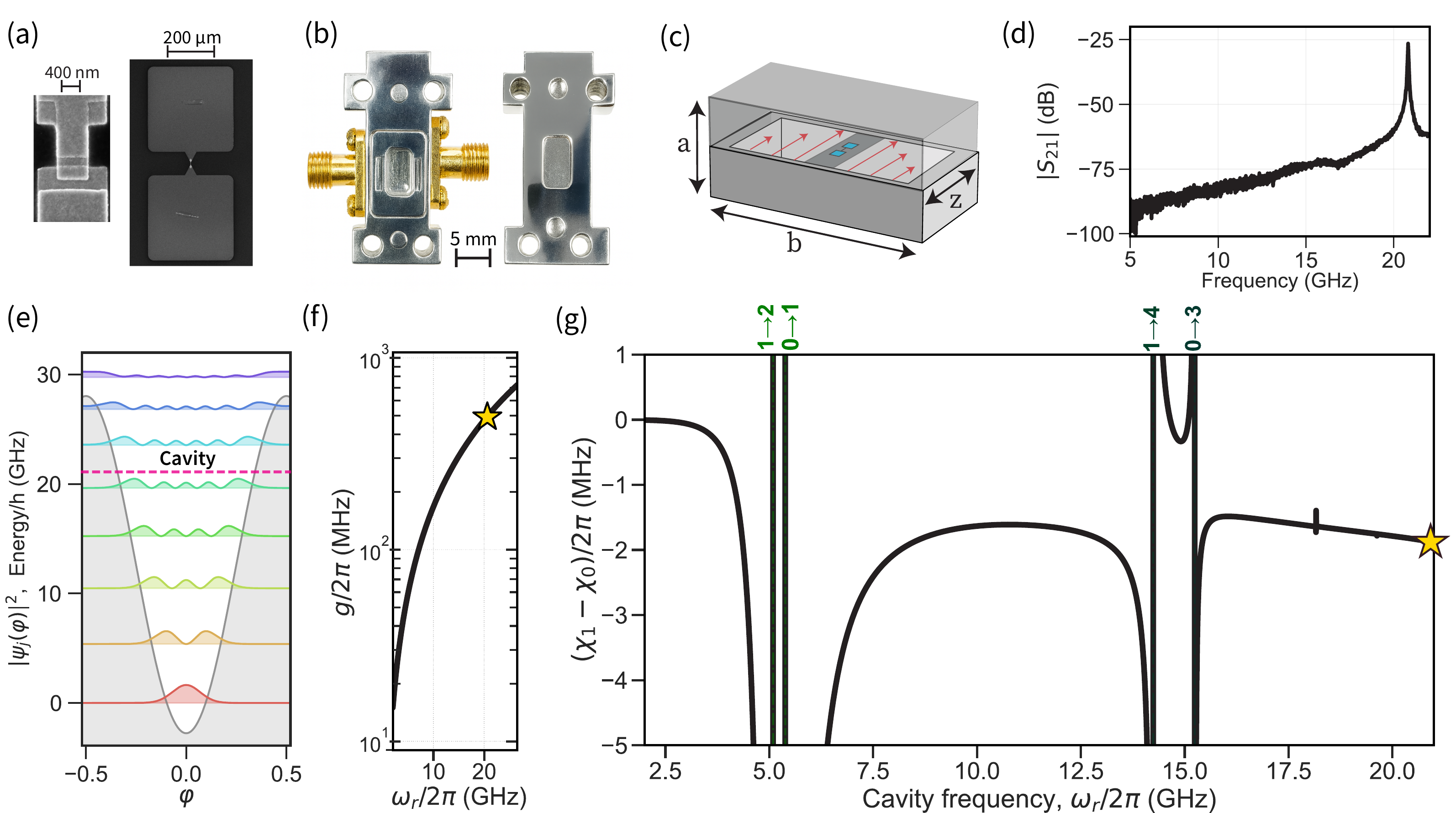}
  \caption{(a) Scanning electron micrographs of the transmon device, showing the AlOx Josephson junction and the Nb capacitor electrodes.  
  (b) Photograph of a rectangular aluminum cavity (made of two pieces) in the two-port transmission measurement configuration. 
  (c) Schematic of Si chip containing a transmon circuit placed inside a 3D rectangular cavity. The electric dipole of the transmon is oriented along the z-dimension.
  (d) Measured room-temperature transmission $S_{21}$ of the cavity containing a qubit chip. (e) Transmon energy levels and wavefunctions inside the Josephson potential. The parameters used in the simulation are from device C. The magenta dashed line indicates the cavity frequency $\omega_r/2\pi$. (f) Theoretical scaling of the qubit-cavity coupling constant $g$ as the cavity dimensions are scaled down. (g) Estimation of the dependence of the dispersive shift $2\chi$ on the cavity frequency (Eqn.~\ref{eq:chi}) as the cavity's dimensions are scaled down. The transmon parameters remain fixed. 
  }
  \label{fig:dispersive_shifts}
\end{figure*}

\section{Dispersive shift in the high cavity frequency limit}
The system under study consists of a fixed-frequency transmon, fabricated on a Si chip, and placed inside a rectangular superconducting cavity (Fig. 1a,b). The only conceptual difference from the original setup reported by H. Paik et al. \cite{paik2011observation} is that the cavity's longitudinal dimensions were scaled down to increase the fundamental mode frequency. The transmon device parameters remain essentially unchanged. The system can be modeled by the following phenomenological Hamiltonian: \begin{equation} \label{eq:hamiltonian}
   \hat{H} = 4E_{C}\hat{n}^2 - E_J \cos \hat{\varphi} + \hbar \omega_r \hat{a}^\dagger \hat{a}+ \hbar g\times i\left( \hat{a}^\dagger - \hat{a} \right) \hat{n},
\end{equation} where $E_C \approx h\times 250~\rm{MHz}$ is the charging energy, $E_J \gg E_C$ is the Josephson energy, $\omega_r$ is the lowest cavity mode frequency, and $g$ is the qubit-cavity coupling constant. The operators $\hat n$, $\hat \varphi$, and $\hat a$ describe, respectively, the Cooper pair number, the conjugate phase-difference, and the cavity photon annihilation operators. We neglect the offset-charge sensitivity of the transmon's higher levels for the purpose of the qualitative discussion below.

In the limit where the qubit frequency $\omega_{01}$ is far detuned from the cavity frequency $\omega_r$, qubit-cavity interaction becomes dispersive and can be modeled by a simple Hamiltonian \cite{koch2007charge,blais2021circuit}: 
\begin{equation}
\hat H/\hbar \approx \frac12 \omega_{01}\hat\sigma_z + \omega_r\hat a^{\dagger}\hat a + \chi\sigma_z a^{\dagger}a.\\
\end{equation}
The parameter scaling of the dispersive shift $\chi$ in the large detuning limit is known as $\chi \sim -E_C(g/(\omega_r - \omega_{01}))^2$. We note that here the parameters $\omega_{01}$ and $\omega_r$ absorb small Lamb-like shifts which we ignore for the purpose of the present work. At a first glance, increasing the detuning $\omega_r - \omega_{01}$ ten-fold from the typical $1-2~\rm{GHz}$ range (say, $5~\rm{GHz}$ qubit and $7~\rm{GHz}$ cavity) to about $15~\rm{GHz}$ (say, $5~\rm{GHz}$ qubit and $20~\rm{GHz}$ cavity) would result in a dramatic suppression of $\chi$ by a factor of about 100. However, such a reasoning does not take into account the fact that increasing the cavity frequency, without modifying the qubit circuit, can lead to a larger value of $g$. Can this increase of $g$ compensate for the higher detuning, such that we keep the desired magnitude of the dispersive interaction $\chi$?

Let us consider a textbook example of a rectangular cavity with the two larger dimensions $a, b$ and a smaller dimension $z$ (Fig.~1c). The fundamental mode has the electric field pointing along $z$ and the transmon's antenna is oriented accordingly for the maximal coupling. In the limit $z\ll a,b$, the mode frequency is given by $\omega_r = c\sqrt{(\pi/a)^2 + (\pi/b)^2}$ ($c$ being the speed of light), so scaling down the cavity's dimensions uniformly leads to the following scaling of the mode volume $V$ with mode frequency
$\omega_r$: $V \propto 1/\omega_r^2$. The RMS electric field in the cavity $E_{\rm{RMS}}$ obeys the general relation $V\times E_{\rm{RMS}}^2 \propto \hbar\omega_r$, that is $E_{\rm{RMS}}^2 \propto \omega_r^3$. Since the qubit circuit is fixed, its electric dipole is fixed as well, and the coupling constant $g$ changes solely because of the change in $E_{\rm{RMS}}$. We thus arrive at a favorable frequency scaling of the coupling constant $g$ of a rectangular cavity:
\begin{equation}
    \label{eqn:g}
    g^2 \propto \omega_r^{3} .
\end{equation}
The above scaling should hold until the cavity becomes so small that the transmon circuit (along with the chip it is fabricated on) modifies substantially the cavity mode. The scaling should also hold approximately even if we keep the $z$-dimension of the cavity fixed, until it remains the smallest of the three dimensions. We thus arrive at a somewhat counterintuitive conclusion that the cubic increase in the coupling constant $g^2$ in a higher frequency 3D cavity, according to Eq.~\ref{eqn:g}, can overcompensate the quadratic suppression of $\chi$ in the large detuning limit. For a cavity in the form of a transmission line resonator, the same reasoning produces a slightly less favorable scaling, $g_{1D}^2 \propto \omega_r^2$, in which case the increase of $g$ would still be sufficient to exactly compensate the effect of larger detuning.

Rigorous calculation of $\chi$ generally requires electromagnetic modeling of the impedance seen by the transmon junction in the vicinity of the cavity frequency~\cite{nigg2012black,minev2021energy}. Such analyses are beyond the scope of our initial work. However, a good estimate of how the value of $\chi$ evolves as one raises the cavity frequency can be obtained from the second-order perturbation theory commonly used to model dispersive shifts in fluxonium qubits~\cite{zhu2013circuit}: 
\begin{equation}\label{eq:chi}
2\chi = \chi_{1}-\chi_{0},~
\chi_{j}=2g^2(\omega_r)\sum_{i\neq j} \frac{\omega_{ij}|\bra{i}\hat{n}\ket{j}|^{2} }{\omega_{ij}^2-\omega_{r}^2},
\end{equation}
where ${j = 0,1}$ and the sum is taken over transitions to all levels in the transmon with $g^2(\omega_r) = g_0^2(\omega_r/\omega_0)^3$. The reference value of the coupling constant $g_0$, corresponding to the reference cavity frequency $\omega_0$, should be chosen to fix the overall scale for $\chi$. An example calculation of $2\chi$ vs. cavity frequency for the parameters of device C in Table \ref{tab:params} is shown in Fig.~1g. One observes that, aside from accidental resonance intersections, the dispersive shift $\chi$ in the high frequency limit is essentially frequency independent, comparable in magnitude to the familiar case of the $7~\rm{GHz}$ cavity, and even has a small growing trend. 

It is instructive to compare our version of the high qubit-cavity detuning regime to that recently described in Ref.~\cite{kurilovich2025high}. There, the detuning was largely due to lowering the transmon frequency rather than increasing the cavity frequency. Therefore, there is no automatic increase in the coupling constant. To keep the value of $\chi$ under control, one must find another way to increase $g$, and the only remaining option is to significantly increase the coupling capacitance between the transmon and the resonator, almost blending the two physical circuits. On the other hand, the maximal ratio of detuning to qubit frequency in our system is only about 4, much larger than in conventional cQED~\cite{dumas2024measurement}, but still significantly below the value of 12 achieved in Ref.~\cite{kurilovich2025high}. Operating the cavity at 24~GHz (the practical edge of the K-band) and slowing down transmons to 4 GHz would further raise the frequency ratio to 6 without modifying our arguments.

\begin{figure}[ht]
  \centering
  \includegraphics[width=7.5cm]{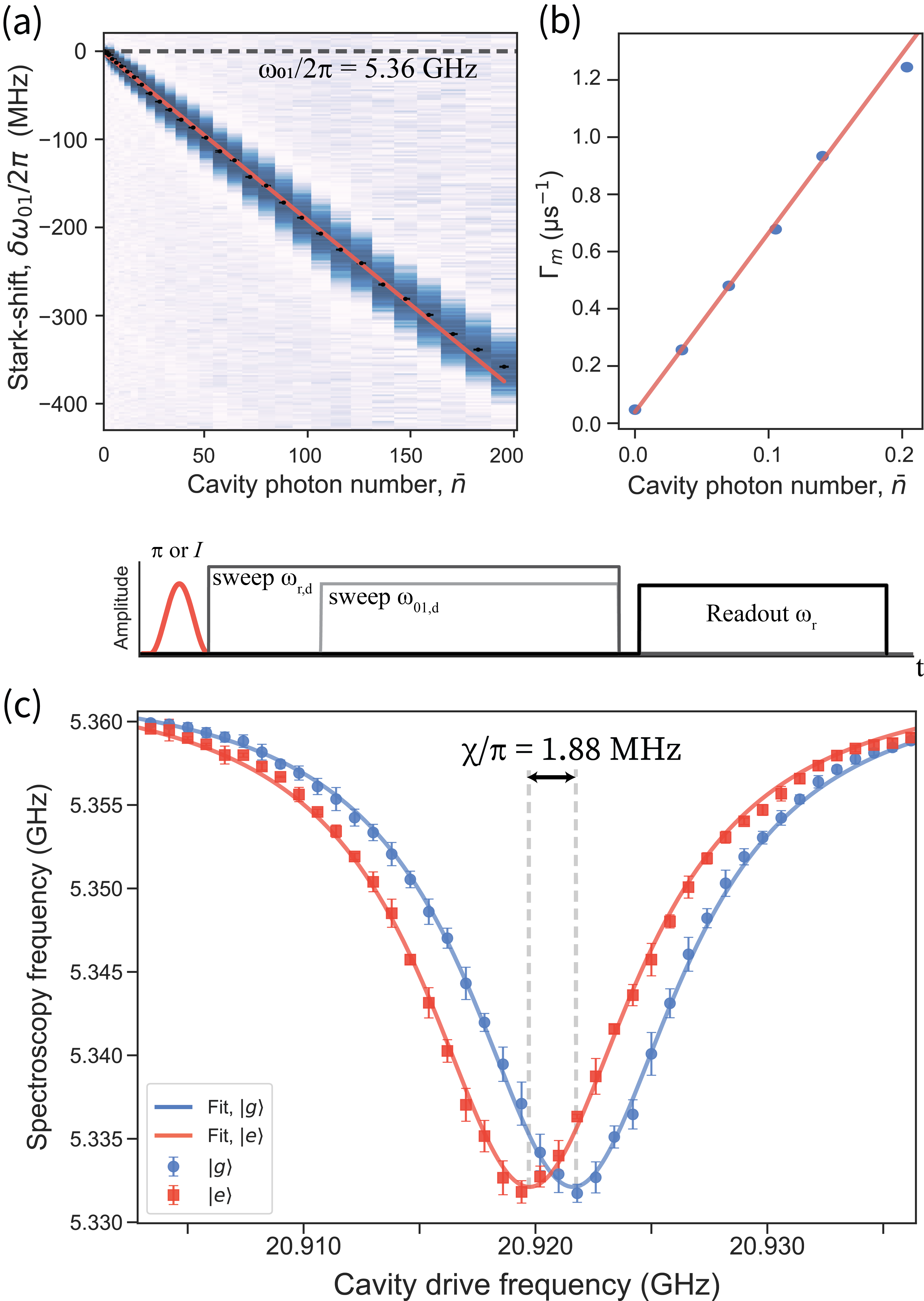}
  \caption{(a) The pulsed ac Stark shift calibration of cavity drive power. The qubit frequency shift, $\delta \omega_{01}$, is linear with cavity photon number and can be used less ambiguously to indicate the relative readout power.  (b) The measurement-induced dephasing rate ($\Gamma_m$) for low cavity photon numbers. The dependence of $\Gamma_m$ on $\bar{n}$ is linear, with the fit values of $\chi$ and $\kappa$ within 10 percent of the values extracted from CKP measurement. (c) The simultaneous fits of the spectroscopy drive, $\omega_{01,d}^*$, with the qubit initially prepared in $|g\rangle$ or $|e\rangle$. The fits are performed simultaneously using Eqn.\eqref{eq:CKP} with $\chi$, $\kappa$, and $\omega_r/2\pi$ as the free parameters. 
}
  \label{fig:CKP}
\end{figure}
 

\section{ Benchmarking 21~GHz cQED}

We chose the cavity frequency in the microwave K-band, at about 21~\rm{GHz}. This frequency choice is a reasonable compromise between implementing a significant frequency increase without running into technical issues, such as high cost and poor availability of appropriate microwave components. Another recent consideration against going too high in frequency is the possibility of quasiparticle generation due to the up-conversion of drive photons to above the gap of superconducting aluminum \cite{chowdhury2025theory}. We measure the cavity in the transmission configuration using a pair of asymmetrically coupled ports. An example of the cavity transmission at room temperature (with the chip inside) is shown in Fig.~1d.
The details of the measurement setup are described in Appendix A while the details of the cavity design are described in Appendix B. We measured four transmon devices labeled A, B, C, D in two measurement lines, with all the extracted parameters shown in Table \ref{tab:params}.

\subsection{Spectroscopy}

The goal of this section is to accurately measure the values of the dispersive shift $\chi$ and the total cavity linewidth $\kappa$, as well as to calibrate the cavity drive in terms of the qubit's ac Stark-shift $\delta \omega_{01}$ and the mean cavity photon occupation $\bar n$ (Fig.~2a) \cite{Schuster2005acStarkShift}. The usual one-tone spectroscopy generally becomes more challenging in our home-made $21~\rm{GHz}$ setup due to the presence of additional up-/down-converting mixers (see Appendix A) and a higher sensitivity to impedance mismatches in the lines. We turn to the ``CKP" technique~\cite{sank2025system, bothara2025high} to accurately extract the values of $\chi$ and $\kappa$. The measurement protocol is the following, after preparing the qubit in either $\lvert 0 \rangle$ or $\lvert 1 \rangle$, a drive at a variable frequency $\omega_{r,d}$ close to the cavity frequency is applied, for a duration longer than $10/\kappa$ and results in shifting the qubit frequency via the ac Stark shift \(\delta\omega_{01}=2\chi\,\bar n\). While the $\omega_{r,d}$ tone is still on, another variable frequency pulse is applied, $\omega_{01,d}$ near the bare qubit frequency. This second tone flips the qubit state if $\omega_{01,d} \approx \omega_{01,d}^{\ast} = \omega_{01} + \delta \omega_{01}$. Both tones are then shut off, allowing the cavity to ring down, and the state of the qubit is measured using a fixed-frequency, fixed-power readout pulse. The frequencies $\omega_{r,d}$ and $\omega_{01,d}$ are swept, essentially implementing two tone spectroscopy of the qubit while simultaneously applying different $\omega_{r,d}$. Depending on initial qubit state, the cavity's resonance frequency shifts up or down by $\chi$ such that the ac Stark shift of the qubit is different at the same value of $\omega_{r,d}$. Each trace is then fit to a single Lorentzian such that $\omega_{01,d}^{\ast}$ is extracted. The extracted, shifted qubit frequency $\omega_{01,d}^{\ast}$ follows the function \cite{sank2025system}: 
\begin{equation}
\label{eq:CKP}
\begin{split}
\omega_{01,d}^\ast &= \omega_{01} + 2 \chi \, \bar n_{g/e}(\omega_d) \\
&= \omega_{01} + 
   \frac{2 \chi \kappa \lvert A \rvert^2}
   {\big(\tfrac{\kappa}{2}\big)^2 + [\omega_{r,d} - (\omega_{r} \mp \chi)]^2} \, 
\end{split}
\end{equation} where A is the drive amplitude of the variable frequency $\omega_{r,d}$. Finally, we fit the two measurements jointly to equation \eqref{eq:CKP} and extract \(\chi\) and \(\kappa\)  (see Fig. 2c). 

To verify the extracted dispersive quantities, the measurement-induced dephasing for $\bar{n}<1$ is fit to the expression $\Gamma_m = \frac{2\bar{n}\kappa \chi^2}{\chi^2 + (\kappa/2)^2}$ \cite{gambetta2006qubit}. This measurement demonstrates the scaling of $\Gamma_m$ at low photon numbers and is dependent on $\chi$ and $\kappa$. The measurement in Fig. 2b fits the dephasing rate as a function of photon number and is accurate to the parameters extracted using the CKP method within 10 percent. 

The extracted values of $\kappa$ for the four devices are between $11-14$~MHz. The variation probably reflects our ability to control the coupling pins of the input and output port connectors. The ports were designed to have asymmetric coupling, with the output port dominating. Having measured the value of $\chi$, we use equation \eqref{eq:chi} to solve for $g$, which comes out in the expected range of about $g/2\pi \approx 500~\textrm{MHz}$. The required values of $E_J$ and $E_C$ for each device were obtained from measuring the frequencies $f_{01}, f_{12}, \text{ and } f_{23}$ and fitting the bare transmon Hamiltonian model at $n_g = 0.25$ offset charge. Our spectroscopic experiment verifies the scaling in Eq.~\ref{eqn:g} and demonstrates that it is indeed possible to consistently achieve a relatively large dispersive interaction $|2\chi| > 1~\rm{MHz}$ with conventional transmons placed inside small-volume high-frequency cavities.

\begin{figure}[ht]
  \centering
  \includegraphics[width=7.5cm]{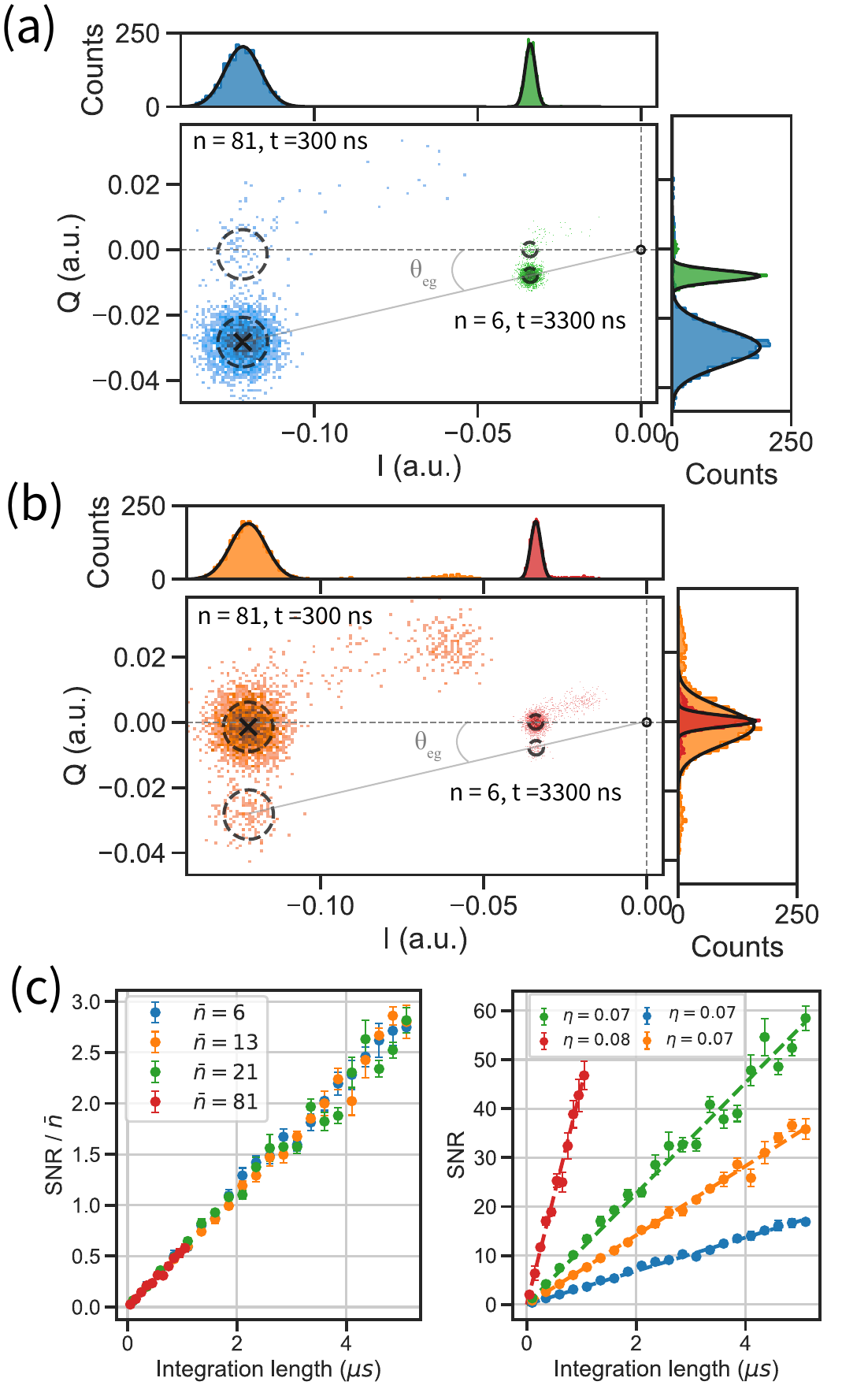}
  \caption{(a) Single-shot-histogram of Device C starting in equilibrium for two values of mean photon number $\bar n$ and integration time $\tau$.
  (b) Same as (a) but with a $\pi_{ge}$-pulse applied prior to the measurement. Note the angle $\theta_{eg}$ is independent on the choice of $\bar n$.
  (c) The $\mathrm{SNR}$ per photon vs $\tau$ extracted from the single-shot data. Note the data respects linear dependence of $\mathrm{SNR}$ on both $\tau$ and $\bar n$ in a broad range of values. Furthermore, the efficiency is independent of mean photon number.   
  }
  \label{fig:efficiency}
\end{figure}
 

\subsection{SNR and quantum efficiency}

Single-shot histograms of a qubit prepared in states $\lvert g \rangle$ or $\lvert e \rangle$ ideally form Gaussian clusters in the IQ-plane centered at \(\mu_{g/e}\) with equal variance \(\sigma^2\) \cite{gusenkova2021quantum,swiadek2024enhancing}. In our devices, the single-shot histogram of the qubit in equilibrium reveals some population of the first excited state, which nevertheless does not prevent us from accurately identifying the values of $\mu_g$ and $\sigma_g$. The single-shot histogram of a qubit in equilibrium to which a $\pi_{ge}$-pulse is applied to further reveals some leakage to a highly-excited transmon state. A detailed study of this leakage process is beyond the scope of the present paper, but we show in Appendix \ref{sec:resonances} that it is consistent with the presence of a single-photon resonance between the readout frequency and a transition from the first to sixth excited state for certain values of the offset charge (for device C). Fortunately, the leakage signal and the excited state signal are well separated in the IQ-plane, which allows an accurate determination of $\mu_e$ and $\sigma_e$ by fitting the data to a Gaussian (Fig 3b). In particular, we verify that the blobs have circular shapes and that $\sigma_e \approx \sigma_g = \sigma$. The signal-to-noise ratio ($\mathrm{SNR}$) is conventionally defined from the IQ-plane data as \(\mathrm{SNR}=|\mu_e-\mu_g|^2/(2\sigma^2)\) \cite{sank2025system}. This definition corresponds to the outcome of the input-output theory \cite{hatridge2013quantum}, given by: \begin{equation}\label{SNR_power}\mathrm{SNR}
= 8\,\eta\,\kappa\,\bar n\, \tau \,\sin^2{\big(\theta_{eg}/2\big)} ~,
\end{equation}
where $\tau$ is the integration time, $\theta_{eg}$ is the angle between vectors $\mu_e$ and $\mu_g$ (the centers of the blobs in the IQ plane corresponding to states $|e\rangle$ and $|g\rangle$). The quantity $\eta$ is the quantum efficiency of the measurement line - the quantity of interest in this section -- defined on the scale from 0 (every photon is lost) to 50\% (quantum limit). We check that the $\mathrm{SNR}$ grows linearly with both $\tau$ and $\bar n$ in a large range of both parameters (Fig.~3a,b). We then equate the experimental and theoretical formulations for $\mathrm{SNR}$ and extract the value of the quantum efficiency $\eta$ as the only remaining unknown parameter. 

For devices B and D,  we find $\eta \approx 4\%$, while for devices A and C, we find $\eta \approx 8\%$. The two pairs of devices were measured in two different measurement lines, which were designed to be as identical as possible. The factor of two difference in the efficiency could be caused by a combination of unequal HEMT noise temperatures, slightly different loss in the isolators, variations in the attenuation of the cables and connectors in the readout line, as well as the variation in the asymmetry of the input and output ports of the cavity. The extracted efficiency can be converted to the system noise temperature $T_{\rm{sys}} = \hbar \omega_r/k_B\eta$ \cite{bothara2025high}, and the case $\eta \approx 8\%$ is equivalent to $T_{\rm{sys}}\approx 12~$K. 
Introducing a quantum-limited amplifier with the noise temperature of about $\hbar\omega_r/2 \approx 0.5~$K would considerably improve the measurement efficiency \cite{hao2025wireless}. Yet, we note that observing $\eta \approx 8\%$ without a paramp is rare in conventional cQED setups \cite{bultink2018general}. The lower value of $\eta \approx 4\%$ in the second measurement line suggests that a factor of two variations in the measurement efficiency are to be expected from setup to setup, but probably not more than that. Our benchmark on $\eta$ establishes the compatibility of today's microwave technology with the highest quantum efficiency measurements in cQED.

\begin{figure*}[ht]
  \centering
  \includegraphics[width=16cm]{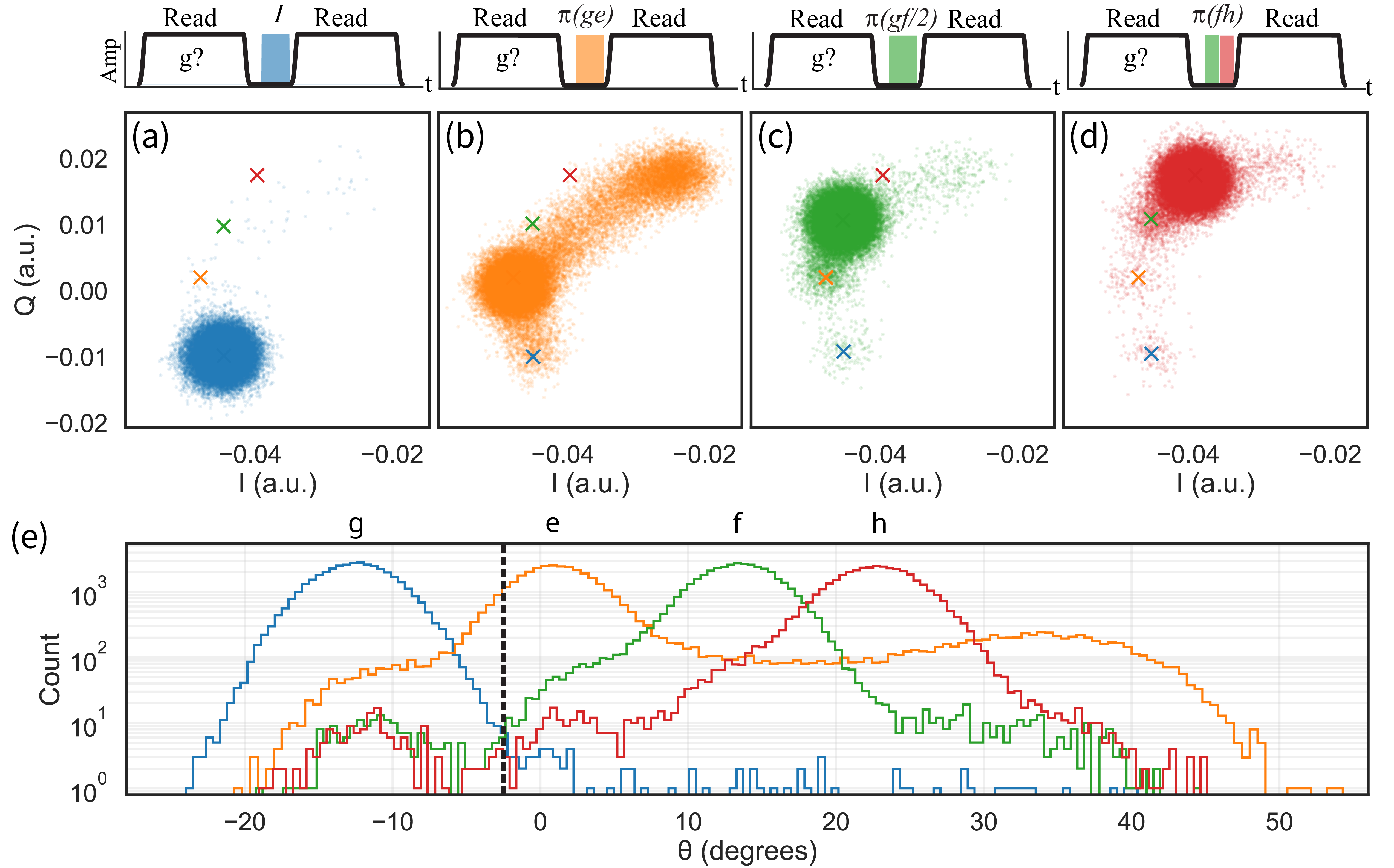}
  \caption{ Results of 4 repeated measurement protocols corresponding to initializing the qubit in states $|g\rangle$ (a); $|e\rangle$ (b); $|f\rangle$ (c); and $|h\rangle$ (d); and observing the measurement outcome.
  The colored ``x'' represents the center of each states blob extracted from a calibration measurement. (d) A histogram of the three data sets above versus angle from the negative x-axis. We use the threshold shown by the black dashed line to separate the outcomes corresponding to $|g\rangle$ state from outcomes corresponding to non-$|g\rangle$ state (in this case $|f\rangle$-state) to define both the initialization error and the assignment error. Notice the suppressed leakage tail in the $|g\rangle$-state histogram in comparison to the higher states. }
  
  \label{fig:QND}
\end{figure*}
 

\subsection{Repeated measurement tests}

The single-shot measurements shown in Fig.~3 suggest that leakage from the ground state is much smaller (if not absent) than leakage from the excited state. Thus, even if the latter effect prevents achieving a proper QND readout, it is valuable to test the ability of the readout to reproduce at least the $|g\rangle$-state outcome in two consecutive measurements. That is, we evaluate the quantity $P(\bar g|g)$, the probability of registering the system not in state $|g\rangle$ in a second measurement, when the first measurement definitely registered the $|g\rangle$ state. The quantity $P(\bar g|g)$ also estimates the error of qubit initialization by post-selection \cite{johnson2012heralded,dassonneville2020fast,bothara2025high} and is useful in its own right; excited states can be similarly initialized by adding the corresponding $\pi$-pulses in between the two readout pulses.

We implemented four repeated measurement protocols shown in Fig.~4. The readout pulse parameters were set to $\bar{n}=14$ ($\delta \omega_{01} = 25$ MHz) and $\tau = 2.5 ~ \mu$s, corresponding to SNR = 14 in Fig.~3. In each of the four protocols, we select only those second readout outcomes for which the first readout passed a conservative ground-state test. The test is such that the readout signal in the IQ-plane lands within one standard deviation of the center of the calibrated position of the $|g\rangle$-state blob. About 40k of such conditional readings are accumulated to reduce the statistical error. We plot the histograms of the angular position of the conditional readout signal in the IQ-plane (Fig.~4e) for a quantitative analysis.

In the first protocol (Fig. 4a), we repeat the readout pulse twice, effectively initializing the qubit in state $|g\rangle$ prior to the measurement; the resulting (blue) histogram has a nearly Gaussian shape, with a very small population of the $|e\rangle$ state and a barely resolvable tail in the higher states angular zone. The deviation from the Gaussian shape is about $10^{-3}$, which provides the first evidence of a suppressed but perhaps not completely absent leakage error from the ground state during the second readout pulse. A simple model of a coherent transmon qubit prepared in the ground state and driven at the cavity frequency does not predict any leakage at the $10^{-3}$ level due to multi-photon resonances at the ac-Stark shift of $\delta \omega_{01} \approx 25~\rm{MHz}$. Further experiments and more sophisticated models are required to explain this effect.

In the second protocol (Fig. 4b), we insert a $\pi_{ge}$-pulse before the second readout pulse to effectively initialize the qubit in state $|e\rangle$ and observe the measurement outcome. The corresponding (orange) histogram has a clear leakage tail extending far away from the $g-e$ angular zone. This leakage is consistent with a trivial first-order resonance between a transition connecting the first to the sixth excited state with the cavity frequency in device C, at a certain value of the offset charge. A detailed account of this effect will be reported elsewhere. The excited state histogram also reveals some counts in the angular zone of the $|g\rangle$-state, the number of which is consistent with the probability $\tau/T_1 \approx 10^{-2}$ ($T_1 \approx 200~\mu$s) for the energy relaxation to the ground state during the readout pulse.

In the two remaining protocols we effectively initialize the second excited state $|f\rangle$ (Fig. 4c) and the third excited state $|h\rangle$ (Fig. 4d) by adding the $\pi$-pulses at $\omega_{gf}/2$ (a two-photon one), and a pair of pulses at $\omega_{gf}/2$ and $\omega_{fh}$, respectively. 
The $|f\rangle$-state histogram (Fig. 4d, green) has a smaller leakage tail, and it also contains counts in the angular zone of the $|e\rangle$- and $|g\rangle$-states. The counts in the $|e\rangle$-zone are consistent with the relaxation of state $|f\rangle$ to state $|e\rangle$ in about $T_1/2 \approx 100~\mu$s, but the counts in the $|g\rangle$-zone should be suppressed by another order of magnitude if they are to be explained by the same relaxation mechanism and assuming the absence of direct decay from $|f\rangle$ to $|g\rangle$. Interestingly, the $|h\rangle$ (red) state histogram has almost the same tail in the $|g\rangle$-zone as the $|f\rangle$ state histogram. At this point we cannot exclude that the nearly equal tails of the $|f\rangle$ and $|h\rangle$ histograms in the $|g\rangle$-zone have the same origin as the smaller tail of the $|g\rangle$ histogram in the $|e\rangle$-zone and are all related to the presence of leakage processes.

We estimate the quantity $P(\bar g|g)$ by choosing the state $|f\rangle$ to play the role of the non-$|g\rangle$ state. We choose a threshold to separate the $|g\rangle$-state and the $|f\rangle$-state  histograms, as shown by the dashed vertical line in Fig. 4d. We sum the counts of the blue histogram to the right side of the threshold to get $P(\bar g|g) \approx 2\times 10^{-3}$. Likewise, we sum the counts in the green histogram to the left of the threshold to estimate the complimentary error $P(g|\bar g)$, that is the probability to register state $|g\rangle$ if the system was prepared in some other state (here $|f\rangle$). We get $P(g|\bar g) \approx 6\times 10^{-3}$. The mean value of the two probabilities defines the readout assignment error for the question ``are you in the ground state?", $\epsilon_{\rm{assignment}} = (P(\bar g| g) + P(g | \bar g))/2 \approx 4\times 10^{-3}$. The ground state initialization error is $\epsilon_{\rm{initialization}} = P(\bar g|g) = 2\times 10^{-3}$. Increasing the readout power increases the $\mathrm{SNR}$ (see Fig.~\ref{fig:efficiency}) but results in a higher value of $P(\bar g|g)$ and also to a higher leakage from the excited states. However, the two errors are already among the lowest reported in the superconducting qubits literature.

\begin{table*}[!bt]
\centering
\resizebox{16cm}{!}{%
\begin{tabular}{
    l  
    S  
    S  
    S  
    S  
    S  
    S  
    S  
    S  
    S  
    S  
    S  
}
\toprule
{} &
{$f_\mathrm{cav}$ [GHz]} &
{$f_{01}$ [GHz]} &
{$E_J/h$ [GHz]} &
{$E_C/h$ [GHz]} &
{$T_1$ [$\mu$s]} &
{$T_{2E}$ [$\mu$s]} &
{$T_{2}^*$ [$\mu$s]} &
{$|\chi|/2\pi$ [MHz]} &
{$\kappa/2\pi$ [MHz]} &
{$g/2\pi$ [MHz]} &
{$\eta$} \\
\midrule
Device A & 20.93 & 6.85 & 26.2 & 0.24 & 150 & 160 & 70*  & 1.43 & 11.46 & 556 & 0.08 \\
Device B & 20.91 & 6.50 & 21.1 & 0.25 & 170 & 240 & 80*  & 0.90 & 13.90  & 461 & 0.04 \\
Device C & 20.92 & 5.36 & 15.4 & 0.26 & 270 & 275 & 100   & 0.94 & 11.28 & 507 & 0.08 \\
Device D & 20.88 & 4.77 & 12.2 & 0.26 & 330 & 280 & 70*   & 0.83 & 14.01  & 510 & 0.04 \\
\bottomrule
\end{tabular}
}%
\caption{Device parameters. The $T_2^*$ data with * indicates double beating in the Ramsey fringe.}
\label{tab:params}
\end{table*}


\subsection{Coherence}

The transmon is measured in the time domain using the standard characterization of energy relaxation ($T_1$),  Hahn-echo coherence ($T_2^E$), and Ramsey interferometry ($T_2^*$). The values of $T_1$ and $T_2$ were measured interleaved over five hours with 200 points each (Fig 5a,b). The values of $T_1$ fluctuate in time but remain north of $100~\mu$s. The origin of these fluctuations remains unknown, but they could be related to the relatively high excited state population of about 6-7\%. This observation invites further improvements for thermalization of the chip and protecting the package from infrared radiation.  The highest reproducible values of $T_1$ correspond to the quality factor $Q>10^7$ (e.g. $T_1 > 350~\mu$s for device D), which is within a factor of two from the best reported transmon relaxation benchmarks~\cite{bland20252d}. There is also a trend that lower frequency devices have longer mean $T_1$ values, which is common to transmons limited by the dielectric loss. The Purcell effect can be safely excluded in the high detuning limit despite a relatively large coupling constant $g/2\pi\approx 500~\rm{MHz}$ for the arguments presented in Ref.~\cite{kurilovich2025high}. Here we note that the same circuit analysis and the same scaling of the Purcell effect relaxation rate $1/T_1(\omega) \sim \kappa(g/\omega_{r})^2\times (\omega/\omega_r)^5$ was already proposed over a decade ago for fluxonium qubits, see Eq.~[B7-B8] of Ref.~\cite{PhysRevB.85.024521}.

The times $T_2^{E}$ are smaller than $2T_1$ but there is a clear correlation in the two times for devices A, B, and C; while device D could be limited by the exponentially small charge dispersion of the qubit transition frequency. The largest observed value of the pure dephasing time $T_{\varphi}$ (defined as $1/T_{\varphi} =1/T_2^{E} - 1/2T_1$) in device D is about $T_{\varphi} \approx 1~$ms. The cavity must have $\bar n \approx 10^{-3}$ residual photon occupation in order to explain such a pure dephasing time. For our cavity such an occupation would be equivalent to a temperature $150~\rm{mK}$, which is unlikely to originate from the lack of thermalization of resistors and attenuators.

\begin{figure}[ht]
  \centering
  \includegraphics[width=8cm]{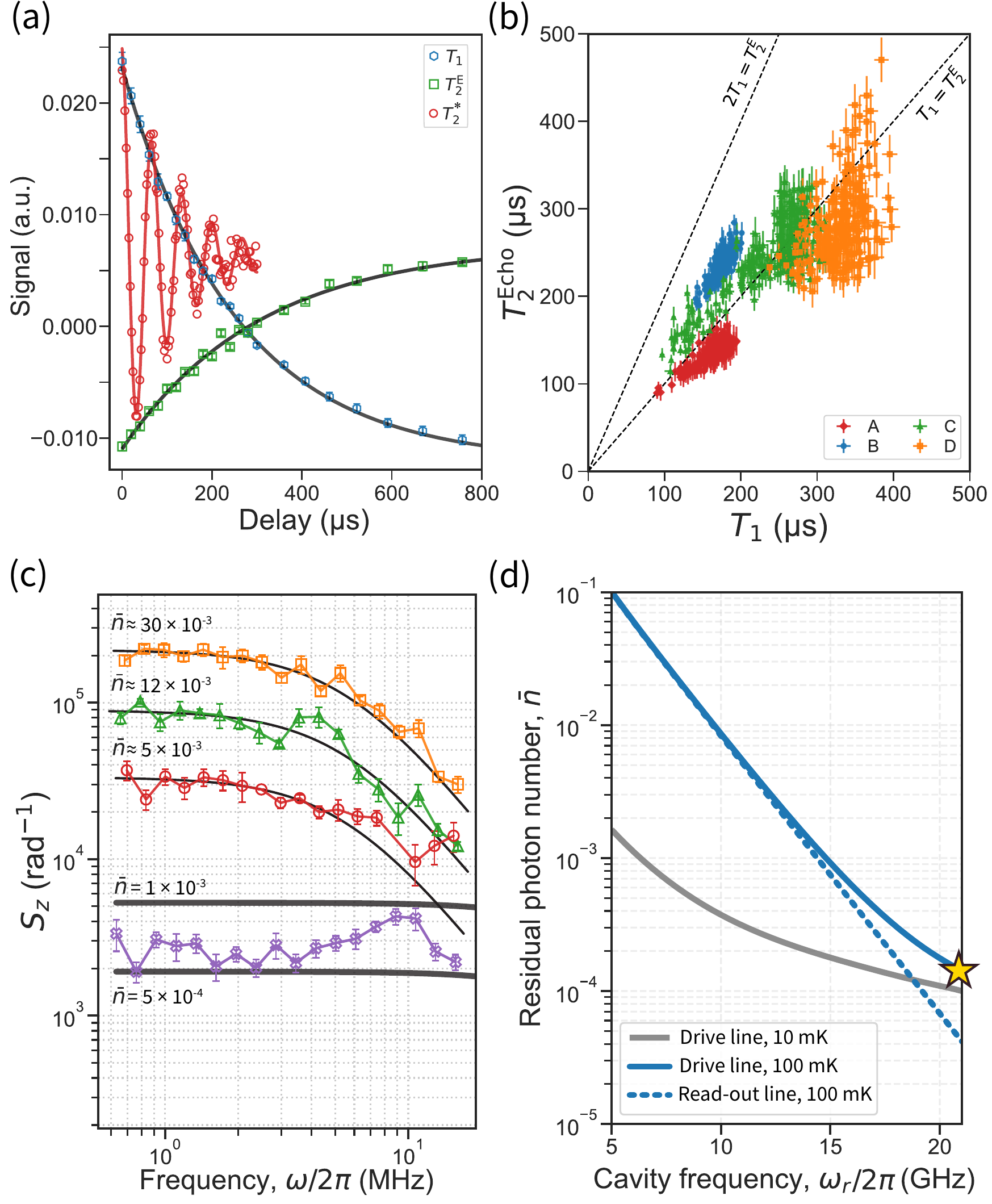}
  \caption{(a) Coherence of device C. The fitted data are from an interleaved measurement with the energy relaxation $T_1 = 250 \pm 3 ~\mu$s,  $T_2^E = 307 \pm 11 ~\mu$s. The Ramsey fringe is fit with $T_2^* = 112 \pm 4 ~\mu$s. (b) Indexed interleaved $T_1$ and $T_2^E$ results for all devices. When $T_1$ increases so does $T_2^E$ indicating the same mechanism limiting energy relaxation could also limit the coherence. The indexed time domain measurements include 200 points per device; fluctuates are seen simulataneously for the interleaved $T_1$ and $T_2^E$ values where the ratio $T_2^E/2T_1<1$. (c) A spin locking measurement shows residual cavity photons $\bar{n} < 10^{-3}$ (purple data). A coherent cavity tone was applied for the other data sets to demonstrate the validity of the experiment.   (d) From the attenuation of the drive line starting at 4 K, there is no significant change to the residual cavity photon number for a 21 GHz cavity if the last attenuator is at 10 mK (gray line) or at 100 mK (blue line). Likewise if the readout line (circulators) were thermalized to 100 mK this would not be the limit of measuring with a 21 GHz cavity (yellow star). }
   \label{fig:coherence}
\end{figure}
 

We further investigated the origin of pure dephasing using the spin locking noise spectroscopy technique. Such a measurement probes the spectral density of the noise of the qubit frequency by inserting a variable amplitude Rabi-drive pulse into a Ramsey sequence \cite{yan2013rotating,yan2016flux}. The readout signal as a function of time decays exponentially with a characteristic decay $T_{\rho}$, related to the noise spectral density $S_z$ at the Rabi frequency $\Omega$, $S_z(\Omega) = 2(1/T_{\rho} - 1/2T_1)$. The presence of residual photons in the cavity with a mean value $\bar n$ corresponds to $S_z(\Omega) =  \big(\frac{8 \bar{n} \chi^2 \kappa^2 }{\kappa^2 + 4\chi^2}\big)\frac{\kappa}{\omega^2 + \kappa^2}$. The results shown in Fig.~5c bound the residual photon population in device C (purple data) at $\bar n \approx 5\times 10^{-4}$. For a sanity check, we applied a calibrated amplitude coherent cavity tone and repeated the spin-locking measurement at elevated mean photon numbers in the cavity $\bar n = (5-30)\times 10^{-3}$ and found a good agreement with theory \cite{yan2018distinguishing} without adjustable parameters. Higher sensitivity to thermal photons and longer, more stable energy relaxation times would help resolve the lower residual photon occupation expected for such a high-frequency cavity. The present data shows little evidence that the measured pure dephasing up to $1~$ms is limited by this effect.

Due to the elevated frequency of the cavity, the readout is less sensitive to the residual photon population. In Fig. 5d, the residual photon population is calculated using standard cascading attenuation theory \cite{pozar2021microwave}, where we simulate from the 4K stage down to the sample stage and compare the impact of modifying the temperature of the last attenuator mounted to the lowest temperature plate. Expectedly, there is no significant impact whether a 20 db attenuator is thermalized to the 10 mK or to the 100 mK temperature plate when using a cavity frequency of 21 GHz. To be complete, we consider the readout line from the 4K stage down to the circulators thermalized to a 100 mK temperature and find this should attenuate sufficiently to values below the $10^{-4}$ level when using a 21 GHz cavity. The insensitivity to attenuator temperature mounted to the coldest plates can clearly provide an advantage since it is well known that thermalization of any component to 10 mK is a significant challenge.  

Finally, $T_2^*$ in three of the samples show a double-beating pattern with the second beat stably detuned on the order of hundreds of kHz. Despite observing the double beating interference pattern, the values $T_2^* > 70~\mu$s while $T_2^*<T_2^E$. The double beating pattern could indicate the presence of two-level systems coupled dispersively to the transmons \cite{schlor2019correlating} and could potentially be a mechanism limiting our dephasing.

\section{Outlook}

In summary, we found no obstacles to increasing the cavity frequency in cQED by at least three-fold to about 21~GHz, without introducing any modifications to the transmon qubit. The magnitude of the qubit-cavity dispersive interaction, the quantum efficiency of the dispersive measurement, and the coherence time of qubits exposed to the new microwave environment all reach state-of-the-art values in multiple devices and experiments. We chose the simplest aluminum box resonator for our initial demonstration, but our approach is also applicable to planar on-chip resonators, provided that one upgrades existing packaging to the higher-frequency standards.

The possibility of operating cQED in the (let's call it) ``ultra-dispersive" limit using a standard qubit and a high-frequency cavity introduces a novel degree of freedom for designing quantum optics and quantum computing experiments. Of highest priority are the investigations into shutting off the leakage in devices with more properly chosen parameters following the findings of \cite{dumas2024measurement, kurilovich2025high, dai2025spectroscopy, connolly2025full}.   
Also of interest is exploring the limits to the scaling shown in Eqn.~3 with even higher-frequency cavities (but perhaps smaller-size chips) for applications, which may involve optomechanics and other hybrid systems~\cite{kumar2023quantum}. Exploring the high-detuning regime for reducing the errors of two-qubit gate operations in numerous schemes based on the cavity bus couplers or bosonic encoding could prove useful as well. The limit to residual thermal population for high-frequency microwave cavities remains an open question. The sensitivity of noise spectroscopy in our experiment could be enhanced by an order of magnitude by using devices with $\chi \approx \kappa$. The observed high-coherence at base temperature invites exploring the resilience of transmons in high-frequency cavities to a higher-temperature environment without being limited by a trivial photon shot-noise dephasing. 

\section{Acknowledgments}

R.A.M. led the experiments and the data analysis, assisted by T.I; samples were provided by I.A.G.; A.L modeled the readout error; V.E.M. proposed and managed the project. We thank Marco Müller, Chulwoo Ahn, and Hao-Chen Yeh for their assistance in testing fabrication procedures, and Shingo Kono for his helpful insights on transmon fabrication and thermalization. We acknowledge funding from SNSF (200021\_213081), SEFRI (20QU-1\_215912), and ARO QC-S5 (W911-NF23-10093) programs.

 \newpage
\bibliography{sources}

\appendix

 
\begin{figure}[ht]
  \centering
  \label{fig:measurement_line}
  \includegraphics[width=8cm]{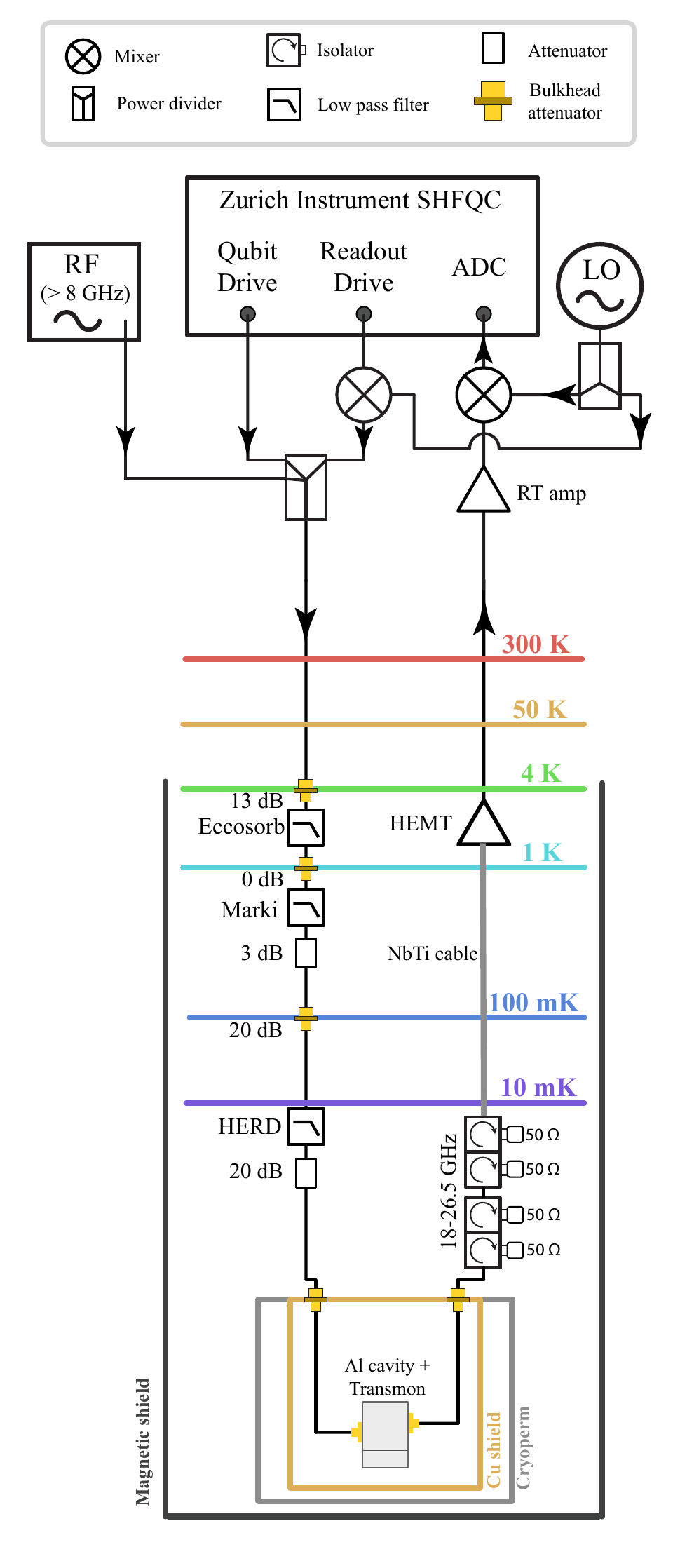}
  \caption{Room temperature electronics connected to the cryogenic measurement line.}
\end{figure}
\section{Experimental setup details}
This experiment used Zurich Instrument SHFQC for all measurements. The RF outputs generated microwave pulses up to 8.5 GHz for the qubit pulses. However, for our high-frequency readout measurements, an 8.0 GHz pulse from the SHFQC is up-converted with an external RF source ($\sim 13$ GHz) using a conventional mixer at room temperature. Both the up-converted readout tone and the qubit drive pulses are combined and sent to the refrigerator through the same microwave line. The returning high frequency pulse out of the fridge, is down-converted at room temperature, using an identical mixer as the input, and is then digitized with the input to the SHFQC at the same frequency of the original readout pulse frequency. (Fig. 6). For experiments using multiple qubit pulses and the CKP measurement, we added another SHFQC drive port and added an auxiliary RF source to apply the cavity-drive pulse at $\omega_{r,d}$.

The cryogenic drive lines are attenuated using commercially available attenuators and filters at each indicated temperature stage of the dilution refrigerator. The filters in the drive line have a cut off starting at around 26 GHz with the last filter being the HERD type \cite{rehammar2023low}. The readout line, starting after the sample, consists of four K-band circulators in series that are thermalized to the 10 mK plate. A single superconducting cable connects the circulators to the 4K HEMT amplifier. All cables and components are of K-type connector style,  since standard SMA connectors are limited to 18 GHz before parasitic effects can significantly degrade the signal.

The Al cavity with the sample is shielded with a copper cylinder, a cryoperm cylinder, and a light-tight stage shield. Furthermore, there is a high-permeability shield enclosing up to the 4K stage and an additional RT magnetic shield comprised of g-iron.

\begin{figure}[t]
  \centering
  \label{fig:cavity_hfss}
  \includegraphics[width=8.5cm]{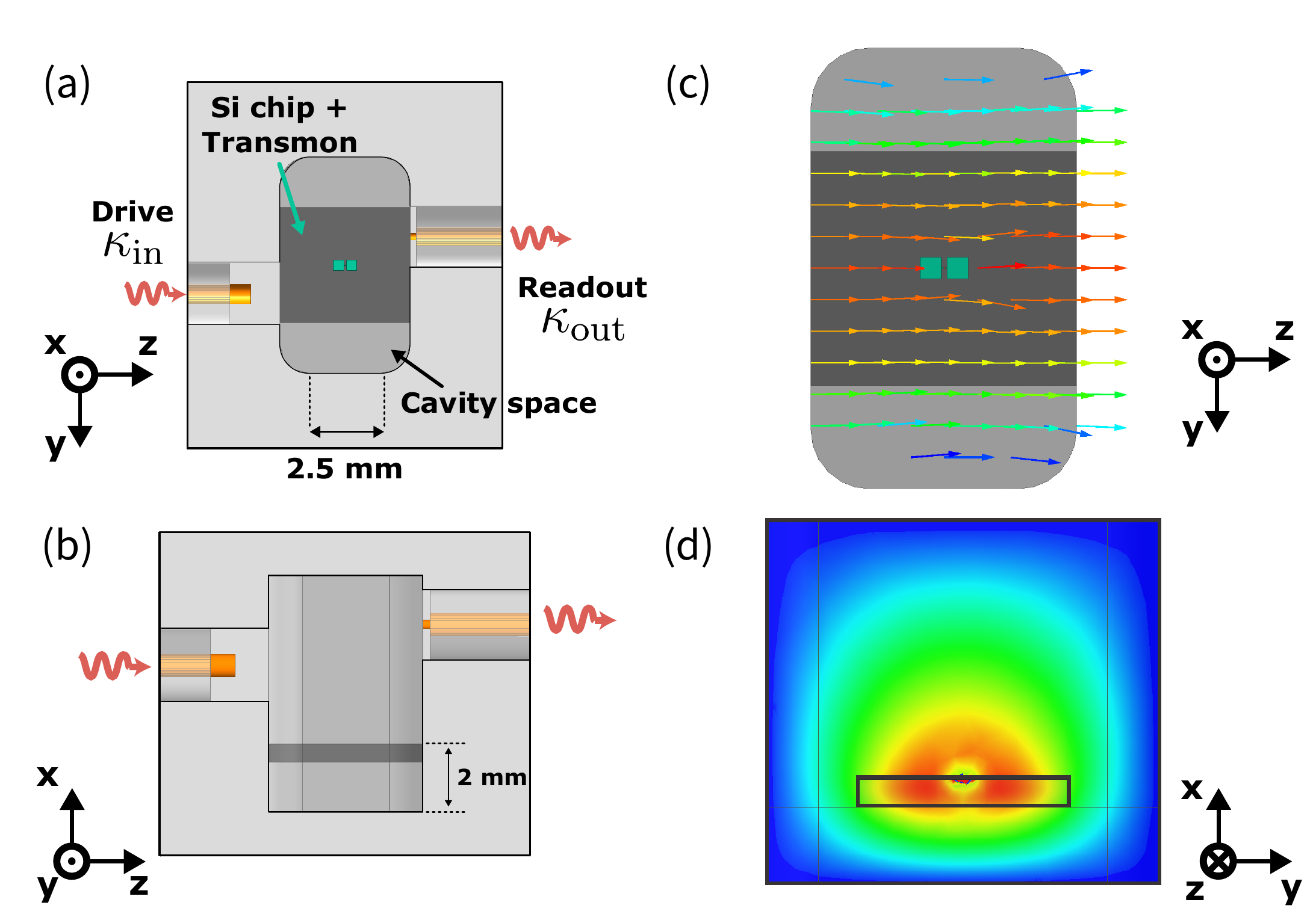}
  \caption{
    (a) Top view and (b) side view of the electromagnetic simulation model of the 21 GHz Al cavity containing the transmon on the Si chip, along with the drive and readout pins. The cavity space dimension is 7.0 mm $\times$ 7.5 mm $\times$ 4.5 mm. The Si chip is placed off-centered in the x axis to make enough space for two pin ports. (c) Simulation of the fundamental TE$_{110}$ mode at 21 GHz. The arrows indicate the electric field at the qubit plane. The color scale represents the field magnitude, which is strongest at the center and weakest at the edges.
    (d) The same fundamental mode on the xy-plane. The field distribution appears slightly off-center due to the chip’s position.
  }
\end{figure}

\section{3D cavity design}
The high-frequency 3D cavity is made from Aluminum 7075-T6. The design is based on simulations using Ansys Electronics HFSS. More specifically, we designed a rectangular waveguide cavity's fundamental mode (TE$_{110}$) at 21 GHz, where the cavity space dimensions were chosen as: 7.0 mm $\times$ 7.5 mm $\times$ 4.5 mm. 

For such a small cavity space, the perturbing effect of the silicon chip as a dielectric cannot be neglected while simulating the resonance frequencies. We found that loading the Si chip in the simulation model could decrease the fundamental mode by up to 10 GHz. The sample chip size is therefore fixed and is placed off-center along the x-axis to allow sufficient space for drive and readout pins. The two pins are placed asymmetrically along the x and y axes to reduce pin-to-pin direct coupling. We use the one closer to the Si chip to drive both the resonator and the qubit, and the other pin is used for readout. The external quality factors for each pin are calibrated at room temperature to create asymmetric coupling, allowing more signal to decay through the readout port. Specifically, the coupling is designed such that $\kappa_{\mathrm{ext}}^{\mathrm{in}}/\kappa_{\mathrm{ext}}^{\mathrm{out}}<1/4$. From the simulations, we find another box mode at 37 GHz which couples to the transmon.


\section{Photon shot noise}
Photon shot noise decoherence in a transmon--cavity system comes from fluctuations in the readout cavity photon number $\hat{n} = a^\dagger a$ caused by its coupling to the measurement line. Thermal photons in the readout cavity randomize the qubit phase since the qubit transition frequency depends on the cavity photon number. The linearized pure dephasing rate from residual cavity photons is:
\begin{equation}
\label{eq:thermal_dephasing}
\Gamma_\phi(\omega_r)
= \frac{4\bar n(\omega_r,T)\,\kappa\,\chi^2}{4\chi^2+\kappa^2},
\end{equation}
where $\bar n(\omega_r,T)=(e^{\hbar\omega_r/k_B T}-1)^{-1}$.

For a fixed $\chi$ and $\kappa$, the dephasing rate sensitivty to the readout frequency is:
${\;
\frac{d\Gamma_\phi}{d\omega_r}
=\frac{\partial \Gamma_\phi}{\partial \bar n}\,
\frac{d\bar n}{d\omega_r}
<0 \;},$
meaning increasing the readout frequency \(\omega_r\) strictly decreases \(\Gamma_\phi\) because \(\bar n(\omega_r,T)\) decreases with \(\omega_r\)  at fixed \(T\). At fixed $T$, increasing the readout frequency $\omega_r$ suppresses the average thermal photon number $\bar{n}$, thereby reducing dephasing. Conversely, even at elevated temperatures, the dephasing rate $\Gamma_\phi$ can be maintained if $\omega_r$ is increased proportionally.

\subsection{Thermal photon number at the sample stage}

We model the input line as a cascade of thermalized attenuation stages 
(300\,K source $\to$ 4\,K $\to$ 800\,mK $\to$ 100\,mK $\to$ 10\,mK). 
For a single stage at temperature $T$, the photon flux from the higher stage is attenuated according to \cite{pozar2021microwave}:
\begin{equation}
    \label{eq:attenuation}
    n_{\mathrm{out}} = L\, n_{\mathrm{in}} + (1-L)\, n_{\mathrm{th}}(f, T),
\end{equation}
where $n_{\mathrm{in}}$ and $n_{\mathrm{out}}$ are the photon numbers entering the attenuator from the higher stage and emitted to the lower stage, respectively. 
$L = 1/10^{A/10}$ is the power attenuation in linear scale, while $A$ is in dB scale. 
Here $n_{\mathrm{th}}(f, T)$ is the thermal (Bose--Einstein) photon number at frequency $f$ and temperature $T$.

The effective photon number at the sample stage is obtained by cascading the contributions of all attenuators and filters in the input line. 
In our simulation, filters introduce an additional frequency-dependent attenuation which was measured with VNA at room temperature and then the attenuation profile plugged back into Eqn. \ref{eq:attenuation} at the given temperature of the stage.

\section{Multiphoton resonances}
\label{sec:resonances}
\begin{figure*}
    \centering
        \includegraphics[width=11cm]{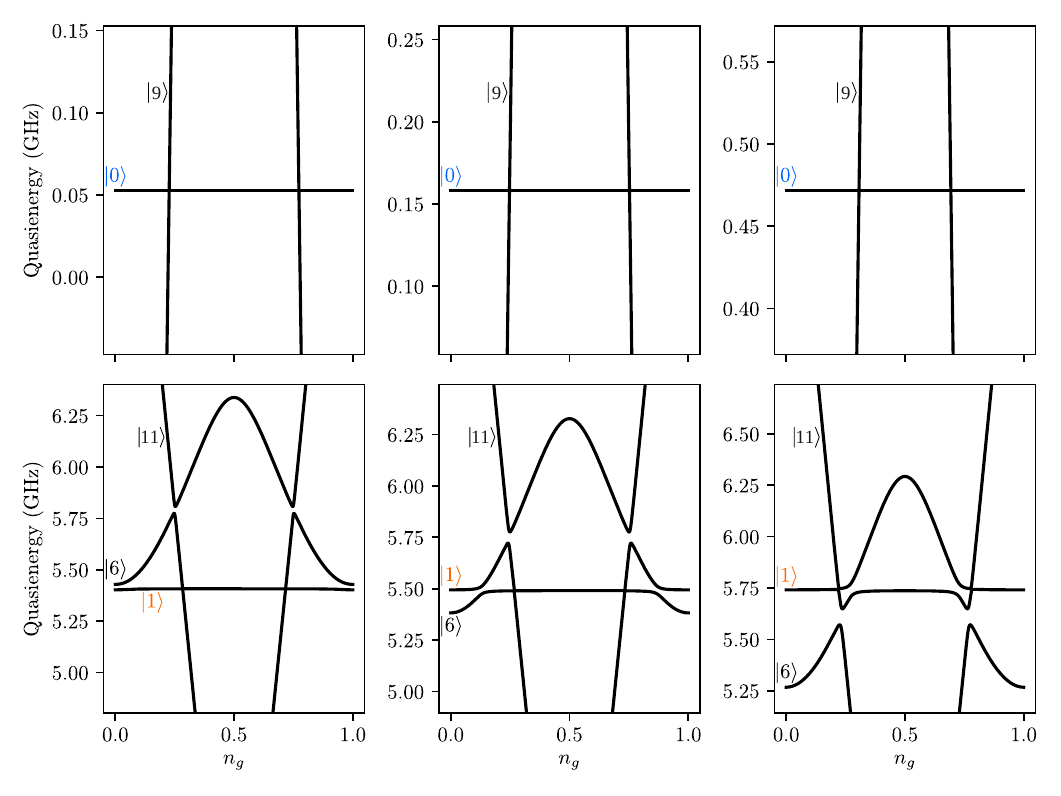}
    \caption{Floquet quasienergies as a function of $n_g$ for device C, at drive amplitudes corresponding to AC Stark shift of the $0-1$ transition of $10,30,90$ sMHz (left to right).
    On the top, the region around state $|0\rangle$ is shown. On the bottom, the region around $|1\rangle$ is shown. Here, anticrossings beween states $|1\rangle$, $|6\rangle$ and $|11\rangle$ are visible, and become more significant at higher drive power. }
    \label{fig:spectrum_ng}
\end{figure*}

To understand the leakage observed in Figs. ~\ref{fig:efficiency}~and~\ref{fig:QND}, we need to model the formation of multiphoton resonances in the case of high-frequency readout. Because the qubit and the cavity are dispersively coupled and the cavity state is well approximated by a coherent state, we model the system as a qubit subject to a classical monochromatic drive, according to the Hamiltonian:\begin{equation}
    \hat{H}_{sc}=4 E_C \hat{n}^2 - E_J \cos(\hat{\varphi}) +2g \sqrt{\bar{n}} \cos(\omega_d t)~.
\end{equation}
The drive is responsible for inducing ac Stark shifts in the qubit energies, and for allowing $m$-photon transitions between qubit states whose energies differ by $m\hbar \omega_d$. Using Floquet theory it is possible to map this problem to a time-independent one, where the energies map to quasienergies defined up to multiples of $\hbar \omega_d$ and multiphoton resonances show up as anticrossings between quasienergy levels.

We plot the quasienergies as a function of $n_g$ for device C in Fig. 8. The plot displays large anticrossings between states $\ket{1}$ and $\ket{6}$, increasing in width with drive power. Conversely, state $\ket{0}$ is not involved in any visible anticrossing. This matches well with the data in Figs. 3,4 and allows us to attribute the leakage from state $\ket{1}$ to a $1\rightarrow6$ single-photon transition.

\end{document}